\documentstyle[12pt,epsfig]{article}
\begin{document}

\newenvironment{QGTSitemize}{\begin{list}{$\bullet$}%
{\setlength{\topsep}{-2.8mm}\setlength{\partopsep}{0.2mm}%
\setlength{\itemsep}{0.2mm}\setlength{\parsep}{0.2mm}}}%
{\end{list}}
\newcounter{enumct}
\newenvironment{QGTSenumerate}{\begin{list}{\arabic{enumct}.}%
{\usecounter{enumct}\setlength{\topsep}{-2.8mm}%
\setlength{\partopsep}{0.2mm}\setlength{\itemsep}{0.2mm}%
\setlength{\parsep}{0.2mm}}}{\end{list}}

                                                                                       
\begin{titlepage}

\vspace*{1.0cm}
\centerline{\large EUROPEAN ORGANIZATION FOR NUCLEAR RESEARCH}

\begin{flushright}
\today
\end{flushright}

\vspace*{1.5cm}

\centerline{{\Large \bf A simple approach to describe hadron production rates}} 
\vspace*{0.3cm}
\centerline{{\Large \bf  in inelastic pp and p$\bar{\mbox{p}}$ collisions}} 
\vskip 0.8in
\centerline{ \large Yi-Jin Pei}
\vspace*{0.3cm}
\centerline{ \large I.~Phys.~Institut der RWTH, D-52074 Aachen, Germany}
\vspace*{0.2cm}
\centerline{ \large  and}
\vspace*{0.2cm}
\centerline{ \large CERN, Ch-1211 Geneva 23, Switzerland}
\vspace*{0.2cm}
\centerline{ \large  (e-mail: Yi-Jin.Pei@cern.ch)}

\vspace*{2cm}

\normalsize
\centerline{{\bf Abstract}} 

\vspace*{0.6cm}

We show that the production rates of light-flavoured mesons and baryons 
in inelastic pp  and p$\bar{\mbox{p}}$ collisions can be described by 
a simple approach used to describe data obtained in $\mbox{e}^{+}\mbox{e}^{-}$ 
annihilation. Based on the idea of string fragmentation, the approach 
describes the production rates of light-flavoured mesons and baryons 
originating from fragmentation in terms of the spin, the binding energy
of the particle, and a strangeness suppression factor. Apart from a
normalization factor and the additional sea quark contribution in inelastic
pp and p$\bar{\mbox{p}}$ collisions, pp, p$\bar{\mbox{p}}$ and 
$\mbox{e}^{+}\mbox{e}^{-}$ data at various centre-of-mass energies are 
described simultaneously.

\vspace*{0.8in}

\begin{center}                                   
\end{center}

\vskip 1.5cm

\vfill
\end{titlepage}

\section{Introduction}

The soft processes of the fragmentation of quarks and gluons into
hadrons cannot be calculated with a perturbative 
approach and instead currently rely on phenomenological models.
The most successful are the string~\cite{string} and the cluster~\cite{cluster}
fragmentation models implemented in the Monte Carlo programs
PYTHIA/JETSET~\cite{JETSET} and HERWIG~\cite{HERWIG}, respectively. 
However, these models require either  a large number of free parameters
in order to reproduce the measured hadron production rates (more
dramatic in the case of PYTHIA/JETSET), or do not give a satisfactory 
description of baryon data in  $\mbox{e}^{+}\mbox{e}^{-}$ annihilation, 
as in the case of HERWIG (a review may be found in Ref.~\cite{sj1,lep2rev}).

In Ref.~\cite{pei} a simple approach 
based on the idea of string fragmentation to describe
hadron production rates in $\mbox{e}^{+}\mbox{e}^{-}$ annihilation is proposed.
We consider that particle production proceeds in two
stages, namely quark pair production in the colour string field and 
successive recombination.
Quark pair production in the colour string field
can be considered as a tunneling process. 
The probability of producing a $q\bar{q}$ pair 
is proportional to exp($-\pi m_{q}^{2}/\kappa$), 
where $m_{q}$ is the (constituent) quark mass, and $\kappa$ the string constant. 
We assume that the probability of quarks recombining to a hadron with the
mass $M_{h}$ is proportional to exp($-E_{bind}/T$), 
where $T$ is the effective temperature in hadronization,
and $E_{bind}=M_{h}- \sum_{i} m_{q_{i}}$ the hadron binding 
energy, which can be ascribed to the colour-magnetic 
hyperfine interaction~\footnote{One could consider 
that the production probability of a hadron with a given
quark content similar to the distribution of the number 
of atoms at different energy levels of the hyperfine splitting 
is determined by the Boltzmann distribution.}~\cite{rosner}. 

The production rates of light-flavoured
mesons and baryons from fragmentation can be described as 
\begin{equation}
 <N> = C \cdot \frac{2J+1}{C_{B}} \cdot (\gamma_{s})^{N_{s}}
       \cdot \mbox{e}^{-\frac{E_{bind}}{T}}, \label{eqrate}
\end{equation}
where $\gamma_{s}=\mbox{exp}[-\pi (m_{s}^{2}-m_{u}^{2})/\kappa ]$ is
the strangeness suppression factor, $N_{s}$ the number of strange quarks
contained in the hadron, and $J$ the spin of the hadron. 
$C$ is an overall normalization factor, 
which increases with the increasing centre-of-mass 
energy (reflecting the rise of multiplicities with increasing energy,
which can be predicted by QCD~\cite{multi}),
and $C_{B}$ is the relative normalization
factor between mesons and baryons (for mesons $C_{B}=1$). 
%
Our approach describes quite well the existing $\mbox{e}^{+}\mbox{e}^{-}$ 
data on hadron production at various centre-of-mass energies. Furthermore,
it can be applied to  heavy flavour production and its predictions there
also agree  well with data.

Multiparticle production in hadron-hadron interactions with low-momentum transfer,
which is still one of the least understood processes  in high-energy physics,
has been studied and compared with particle production in $\mbox{e}^{+}\mbox{e}^{-}$ annihilation
by many authors~\cite{basile,bard,chliap,marek,becatt}. 
It is generally believed  that the mechanism of hadronization is the same in all
high-energy processes, and that all differences in particle composition
can be traced to different initial parton configurations~\footnote{The Lund
string model~\cite{string,JETSET}, for example, is one explicit realization of
this concept.}. It is therefore important to test our approach with
data obtained in  hadron-hadron collisions.

In this paper, we first show some similarities between the hadron production rates 
measured in inelastic pp collisions and in $\mbox{e}^{+}\mbox{e}^{-}$ annihilation. 
Then, taking into account the different initial parton 
configurations and the contribution from sea quarks, we apply our approach
to data obtained in inelastic pp and p$\bar{\mbox{p}}$ collisions.

\section{Data on hadron production in pp and p$\bar{\mbox{p}}$ collisions}
The data on hadron production in inelastic pp and 
p$\bar{\mbox{p}}$ collisions used in this analysis are listed in Tables~1--4.
In some papers, listed in the Tables, pp data are given in terms of production cross
sections instead of production rates.  
To calculate the production rate
of hadrons per inelastic pp collision, we use the value of 
the total inelastic pp cross sections quoted in the papers,
or the value from Ref.~\cite{PDG} at the corresponding centre-of-mass energy 
if inelastic pp  cross sections are not given in the papers. 
When counting the hadron production rates, the decay products of 
$\mbox{K}^{0}_{S}$, $\Lambda$, $\Sigma^{\pm}$, $\Xi^{0,-}$ and
$\Omega^{-}$ (and their antiparticles)  are not included in the papers. 
This is different from the counting procedure used for 
$\mbox{e}^{+}\mbox{e}^{-}$ data.

The most complete set of data on hadron production in inelastic pp collisions
consists of the LEBC-EHS results~\cite{LEBC} at $p_{\mbox{\tiny LAB}}=400$~GeV 
(centre-of-mass energy $\sqrt{s}=27.4$~GeV), and the Fermilab 30-inch bubble
chamber data~\cite{kichimi} at $p_{\mbox{\tiny LAB}}=405$~GeV ($\sqrt{s}=27.6$~GeV).
This data set  can be compared with the LEP data~\cite{QGYPange}, 
the most complete set of data on hadron production 
in $\mbox{e}^{+}\mbox{e}^{-}$ annihilation. 
Although the type of interactions and the initial parton configuration
in pp interactions with low-momentum transfer 
are different from those in $\mbox{e}^{+}\mbox{e}^{-}$ annihilation,
for a direct comparison 
we can however  use hadrons which do not contain the valence quarks (u,d) in the proton,
such as $\bar{\mbox{u}}$s, $\bar{\mbox{d}}$s and s$\bar{\mbox{s}}$ mesons,
and antibaryons. In most of the cases, such hadrons  are either produced from 
fragmentation, or are decay products of other hadrons 
which are  produced from fragmentation~\footnote{The contribution of sea quarks
and antiquarks to hadron production is small (see discussions in the next section)}. 
On the other hand,
the contribution of the primary $q\bar{q}$ pair to the production rate
of light-flavoured hadrons is small at LEP.
It is therefore possible to use the following hadrons to compare directly 
hadron production from fragmentation in inelastic pp collisions and in
$\mbox{e}^{+}\mbox{e}^{-}$ annihilation:
\begin{center}
$
\begin{array}{lclc} 
\mbox{Hadron~~~~} &n_{pp}(27.5\mbox{GeV})/n_{ee}(91\mbox{GeV}) 
& \mbox{~~~~~~~~~~Hadron~~~~}  & n_{pp}(27.5\mbox{GeV})/n_{ee}(91\mbox{GeV}) \\
\mbox{~~}\mbox{K}^{-}        & 0.19\pm 0.01 
& \mbox{~~~~~~~~~~~~}\bar{\mbox{p}}    & 0.13\pm 0.01  \\    
\mbox{~~}\mbox{K}^{*-}       & 0.25\pm 0.04 
& \mbox{~~~~~~~~~~~~}\bar{\Lambda}     & 0.11\pm 0.02  \\   
\mbox{~~}\bar{\mbox{K}}^{*0} & 0.24\pm 0.04 
& \mbox{~~~~~~~~~~~~}\bar{\Delta}^{--} & 0.20\pm 0.13  \\    
\mbox{~~}\phi                & 0.18\pm 0.02 
& \mbox{~~~~~~~~~~~~}\bar{\Sigma}^{\pm}& 0.35\pm 0.13     
\end{array} 
$  
\end{center}   
For all hadrons except  the $\bar{\mbox{p}}$ and $\bar{\Lambda}$, 
the ratio $n_{pp}$/$n_{ee}$ is about the same within errors, 
indicating that  hadron production from fragmentation in inelastic pp collisions
and in $\mbox{e}^{+}\mbox{e}^{-}$ annihilation differs only by a normalization
factor, i.e., the available energy in fragmentation. 
For the $\bar{\mbox{p}}$ and $\bar{\Lambda}$, this ratio is lower than that for other
hadrons, mainly  due to the different counting procedures used for decay products
of the $\Lambda$, $\Sigma^{\pm}$, $\Xi^{0,-}$ and
$\Omega^{-}$ (and their antiparticles) as mentioned above, 
and to the additional contribution to the $\bar{\mbox{p}}$ and $\bar{\Lambda}$
from decays of c and b baryons in  $\mbox{e}^{+}\mbox{e}^{-}$ annihilation.
If these two corrections are taken into account, the ratio $n_{pp}$/$n_{ee}$ 
is then equal to $0.21\pm 0.02$ for the $\bar{\mbox{p}}$~\footnote{For the 
$\bar{\Lambda}$ the uncertainty of the corrections is large since the branching
fractions c(b) baryons $\rightarrow \bar{\Lambda}$ are not well known.}.
From the values of the ratio $n_{pp}$/$n_{ee}$ listed above,  we see that
the number of hadrons produced from fragmentation in inelastic pp collisions 
at $\sqrt{s}\simeq 27.5$~GeV is about 20\% of that in  $\mbox{e}^{+}\mbox{e}^{-}$  
annihilation at $\sqrt{s}=91$~GeV. We will discuss this
in more detail in the next section.

\section{Analysis}
%
We first consider  hadron production in inelastic pp collisions. 
Similar to hadron production in $\mbox{e}^{+}\mbox{e}^{-}$  
annihilation, this process can also be considered to proceed
in four steps: interaction between two initiator partons (valence quarks,
gluons, sea quarks or antiquarks) with two beam remnants left behind,
followed by a parton shower development from the initiator partons, 
and subsequently the transition from partons to hadrons. Finally, 
the unstable hadrons decay according to their branching ratios. 

For a proton the possible scenarios for the initiator parton and beam
remnant are the following:
\begin{enumerate}
 \item If the initiator parton is a valence quark $q_{v}$, the beam remnant
       is simply a diquark composed of the two leftover valence quarks,
       i.e. either a ud or a uu diquark. The initiator parton and
       the diquark are at the two end-points of a string which is then
       hadronized in the usual way~\cite{JETSET}.
 \item If the initiator parton is a sea quark $q_{s}$ or antiquark $\bar{q}_{s}$,
       the beam remnant contains  four quarks: uud$\bar{q}_{s}$
       or uud$q_{s}$. Since the $q_{s}\bar{q}_{s}$ pair, to a first
       approximation, is in a  colour-octet, the subdivision uud$+q_{s}$
       (or uud$+\bar{q}_{s}$) is not allowed, since it would correspond to 
       a colour-singlet $q_{s}\bar{q}_{s}$~\cite{JETSET}. Therefore
       uud will be subdivided into a quark (u or d) and a diquark (ud or uu).
       In this case in the  beam remnant  we have a diquark and two
       single quarks (or a quark and a antiquark), 
       which will then be at the end-points of two strings~\footnote{Here we use 
       the treatment of sea quarks and antiquarks as discussed in Ref.~\cite{ingelmann}, 
       which is different from the conventional way~\cite{JETSET}.}
       and hadronized in the usual way.
       We introduce two free parameters  for the fractions of the sea quarks: 
       $f=f_{u}=f_{d}$ and $x_{s}=f_{s}/f$.
 \item If the initiator parton is a gluon $g$, the beam remnant is a colour-octet
       uud state, which is subdivided into a quark and a diquark. The
       treatment is similar to that for scenario 1.
\end{enumerate}
On average, the individual numbers of the primary quarks  and diquarks  
in an inelastic pp collision are: $N_{\mbox{\tiny u}}=2(\frac{2}{3}+f)$, 
$N_{\mbox{\tiny d}}=2(\frac{1}{3}+f)$, 
$N_{\bar{\mbox{\tiny u}}}=N_{\bar{\mbox{\tiny d}}}=2f$,
$N_{\mbox{\tiny s}}=N_{\bar{\mbox{\tiny s}}}=2x_{s}f$,
$N_{\mbox{\tiny uu}}=2\times \frac{1}{3}$ and $N_{\mbox{\tiny ud}}=2\times \frac{2}{3}$.

Taking into account the numbers of the primary quarks  and diquarks, 
hadron production in inelastic pp collisions can also be described
by our approach discussed in Ref.~\cite{pei}.
For simplicity, as in Ref.~\cite{pei}, we apply our approach analytically to the data.
Instead of using the parton distributions $p(x,Q^{2})$ 
($p=q_{v},q_{s},\bar{q}_{s},g$), for the above three scenarios  we use 
an averaged parameter $C$ in  Eq.~(\ref{eqrate}), which is  
integrated over $x$. 
The total hadron rates are calculated as follows. Firstly we calculate
the number of primary hadrons. Primary hadrons are defined
as those which are not decay products of other hadrons, i.e. as hadrons
originating from fragmentation, or containing a primary quark  or a diquark from the 
p in an inelastic pp collision. 
We use Eq.~(\ref{eqrate}) to calculate the 
number of light-flavoured hadrons produced from fragmentation. 
Heavy quark production from fragmentation
is strongly suppressed due to the term exp($-\pi m_{q}^{2}/\kappa$)
and can therefore be neglected.
For hadrons which contain a primary quark $q$($q=q_{v},q_{s},\bar{q}_{s}$)
or a primary diquark $q^{1}_{v}q^{2}_{v}$, 
we use Eq.~(\ref{eqrate}) to determine their relative ratios, and then obtain
their rates by normalizing the sum of the rates to the average number of
$q$ or $q^{1}_{v}q^{2}_{v}$ in an inelastic pp 
collision. Compared to the production cross section of light-flavoured hadrons,
the charm and bottom production cross sections in pp collisions are 
small~\cite{ammar} and can therefore be neglected in this analysis.
The spin of the diquark is taken into account in the calculation. For example,
the $\Lambda$ contains a spin-0 ud and the $\Sigma^{0}$ a spin-1 ud diquark.
All light-flavoured hadrons up to a mass of 2.5~GeV in the meson and
baryon summary table of Ref.~\cite{PDG} are included in the calculation. 
Finally, we let all the primary hadrons decay according to 
their decay channels and branching ratios given in Ref.~\cite{PDG}.
The decay chain stops when $\mu $, $\pi$, $\mbox{K}^{\pm}$, $\mbox{K}^{0}_{L}$,
$\mbox{K}^{0}_{S}$, $\Lambda$, $\Sigma^{\pm}$, $\Xi^{0,-}$,
$\Omega^{-}$ (or their antiparticles) or stable particles are reached.

In the fit to pp data at $\sqrt{s}=27.4$--$27.6$~GeV, we choose $C$, $f$ and $x_{s}$
as free parameters. For the other parameters in Eq.~(\ref{eqrate}) we use the result
of the fit to $\mbox{e}^{+}\mbox{e}^{-}$ data~\cite{pei}: $\gamma_{s}=0.29\pm 0.02$,
$\Delta m = 0.161\pm 0.024$~(GeV), $T=0.298\pm 0.015$~(GeV) and $C_{B}=11.0\pm 0.9$.
In the fit these parameters are set  to their central values
since otherwise there would be too many free parameters which
are highly correlated with each other, leading to unstable fit results.
The value of $m_{s}$ is set to 
0.5~GeV~\footnote{Since exp$(-E_{bind}/T)=\mbox{exp}(\sum_{i} m_{u}/T)
\cdot \mbox{exp}\{-[M_{h}-\sum_{i} (m_{q_{i}}-m_{u}) ]/T\}$,  the factor
$\mbox{exp}(\sum_{i} m_{u}/T)$ can be absorbed in $C$ and $C_{B}$.
The error function of the fit is mainly sensitive to the change in the 
mass difference  $m_{s}-m_{u}$.}. 
The fit is performed by minimizing the error function
\begin{equation}
 \chi^{2} = \sum_{i} (N^{calc}_{i} - N^{meas}_{i})^{2}/ (\Delta N^{meas}_{i})^{2}\mbox{~,}
  \label{eqchi}
\end{equation}
where $N^{calc}_{i}$ is the calculated rate, $N^{meas}_{i}$ and 
$\Delta N^{meas}_{i}$ the measured rate and its error, respectively.

The fit results are listed in Table~5 and shown in Fig.~1. The value
of the normalization factor $C$ is equal to $0.042\pm 0.007$, where the error
also includes the uncertainty on  $\gamma_{s}$, $\Delta m $, $T$ and $C_{B}$ which
are set  to their central values in the fit. 
The ratio of this value and the value of $C$ obtained from the fit to 
$\mbox{e}^{+}\mbox{e}^{-}$ data at 
$\sqrt{s}=91$~GeV~\cite{pei} is $(19\pm 3)$\% (common systematic errors have been taken 
into account), which is in a good agreement
with the value of $n_{pp}/n_{ee}$ discussed in the previous section.
From the fit, the probability that the initiator parton is a u (or d) sea quark
is determined to be $f=0.044\pm 0.010$. The strange sea content is determined
to be $x_{s}=f_{s}/f=0.51\pm 0.22$, which is consistent with the parametrization
used in Ref.~\cite{martin}. 

If we  also choose the parameter $T$  as a free
parameter in the fit, we obtain $T=0.271\pm 0.010$~GeV, which is in agreement
with the values of $T$ obtained from the fits to 
$\mbox{e}^{+}\mbox{e}^{-}$ data~\cite{pei}. This indicates that the value of $T$ is
universal in different types of interactions.

As can be seen from Table~2 and Fig.~1, for most of the hadrons the calculated rate agrees well
with the measurement (within two standard deviations). 
Only at three data points, namely $\pi^{+}$, $\rho^0$ and $\Sigma^{*-}$, is the difference
between the calculated and measured rates  larger than three standard deviations.
These three data points contribute more than  half of the  
$\chi^{2}$ value of the fit (49 out of the total $\chi^{2}$ of 85 for 33 data points,
fitted with three free parameters). 
Another major reason  for the large $\chi^{2}$ 
value is because the uncertainty of the calculated rates due to  
uncertainties on the mass,  branching ratios and decay modes of the 
resonance states is not taken into account in the fit. If, for example, we introduce
a 5\% error on the number of secondary  particles as the uncertainty of the 
calculated rates ($\Delta N^{calc}_{i} = 0.05*(1-f^{prim}_{i})*N^{calc}_{i}$,
where $f^{prim}_{i}$ is the primary fraction of hadron $i$)
and add it to the term $\Delta N^{meas}_{i}$ in Eq.~(\ref{eqchi}), 
the $\chi^{2}/dof$ value of the fit is then reduced from 85/30 to 61/30, 
while the fit results remain essentially unchanged. 
Comparing the fit 
to almost the same data set  with the same number of free parameters in Ref.~\cite{becatt},
the $\chi^{2}/dof$ value of our fit is much lower~\footnote{Note in Ref.~\cite{becatt}
a $\Delta N^{calc}_{i}$ term  has been  taken into account in the fit. In some cases,
for example, for $\rho^{0}$ and $\Delta^{++}$, $\Delta N^{calc}_{i}$ is much larger than 
$\Delta N^{meas}_{i}$.} and our calculated rates are closer to the measured rates.

In the fit to pp and p$\bar{\mbox{p}}$ data at the other centre-of-mass energies
we choose only $C$ as a free parameter because of the small number of  data points 
available at some energies.
The parameters $f$ and $x_{s}$ are set at the values obtained from the fit
to pp data at $\sqrt{s}=27.4$--$27.6$~GeV. 
In the pp data set at $\sqrt{s}=52.5$--$53.0$~GeV, f$_2(1270)$, K$^{*\pm}_{2}(1430)$ and
$\bar{\mbox{K}}^{*\pm}_{2}(1430)$ are not included in the fit as
the results from Ref.~\cite{drijard} and \cite{bohm} differ significantly.
The treatment of initiator partons and beam remnants in p$\bar{\mbox{p}}$ collisions
is similar to that in pp collisions, with one of the two initiator partons and beam remnants
replaced by their antiparticles.
The fit results are listed in Tables~1, 3, 4 and 5.
One can see that the calculated rates agree well with the measurements.

The fraction of primary hadrons is also given in Tables~1--4.
Owning to different initial parton configurations, the fraction values in Tables~1--4 
are different from the corresponding values for hadrons produced in 
$\mbox{e}^{+}\mbox{e}^{-}$ annihilation at $\sqrt{s}=91$~GeV (see the last column
in Table~2): 
\begin{itemize}
 \item For mesons, and in inelastic pp collisions also antibaryons, the fraction value 
       for pp (p$\bar{\mbox{p}}$) data is in general slightly higher 
       than that for $\mbox{e}^{+}\mbox{e}^{-}$ data, 
       mainly due to the fact that the amount of c and b hadrons 
       produced in pp (p$\bar{\mbox{p}}$) collisions is negligible. 
       In contrast, in $\mbox{e}^{+}\mbox{e}^{-}$ annihilation at $\sqrt{s}=91$~GeV
       about 40\% of the events contain c or b hadrons which decay to light-flavoured 
       hadrons, leading to a lower value of the primary fraction of light-flavoured hadrons.
 \item For baryons which contain a uu ($\bar{\mbox{u}}\bar{\mbox{u}}$) or 
       ud ($\bar{\mbox{u}}\bar{\mbox{d}}$) diquark, the fraction value 
       for pp (p$\bar{\mbox{p}}$) data  is much higher
       than that for $\mbox{e}^{+}\mbox{e}^{-}$ data 
       as  a uu ($\bar{\mbox{u}}\bar{\mbox{u}}$) or 
       ud ($\bar{\mbox{u}}\bar{\mbox{d}}$) diquark already exists in the
       initial configuration of inelastic pp (p$\bar{\mbox{p}}$) collisions.
       In contrast, for ddd ($\bar{\mbox{d}}\bar{\mbox{d}}\bar{\mbox{d}}$) or 
       dds ($\bar{\mbox{d}}\bar{\mbox{d}}\bar{\mbox{s}}$) types of baryons, for example,
       the $\Sigma^{-}$, the fraction value can be lower than that for 
       $\mbox{e}^{+}\mbox{e}^{-}$ data as such baryons cannot be
       produced directly from the diquark in the p ($\bar{\mbox{p}}$), but 
       can be decay 
       products of the other baryons which are produced directly from the diquark 
       in the p ($\bar{\mbox{p}}$).
\end{itemize}

\section{Conclusions}

We have shown that the production rates of light-flavoured mesons and baryons 
in inelastic pp and p$\bar{\mbox{p}}$ collisions are described by a simple approach
used to describe  $\mbox{e}^{+}\mbox{e}^{-}$  data. 
Taking into account the different initial parton configurations and 
the additional sea quark contribution in  inelastic pp and
p$\bar{\mbox{p}}$ collisions, pp, p$\bar{\mbox{p}}$ and $\mbox{e}^{+}\mbox{e}^{-}$ 
data on the  production of light-flavoured hadrons
at various centre-of-mass energies are described simultaneously
(apart from a normalization factor, which reflects the rise of multiplicities 
with increasing energy). Moreover, as shown in Ref.~\cite{pei}, data on heavy
flavour production in $\mbox{e}^{+}\mbox{e}^{-}$ annihilation are also 
described by our approach. 
All this shows that our approach can provide a universal description of
hadron production, irrespective of the type of the interaction and the initial 
parton configuration in various interactions.

\section*{Acknowledgements}

We are grateful to  F.~Becattini, T.~Hebbeker, R.J.~Hemingway and  
T.~Sj\"ostrand for useful discussions. We also wish to thank 
J.~Navarria for a careful reading of the draft of this paper.

                                                       
\newpage

\begin{table}[th]
\begin{center}
\begin{tabular}{lcccc} \hline   
 Hadron      &  Rate     &  Rate      & Primary Fraction & References  \\   
             &  Measured & Calculated & Calculated       &  \\ \hline   
\multicolumn{5}{l} {$\sqrt{s} =19.4$--$19.7$~GeV ($\sigma_{inel}=32.1$~mb~\cite{barish})}      \\ \hline    
      Charged          & $7.68\pm0.06$    &    7.53   &                  & \cite{barish,allday} \\
   Neg. charged        & $2.85\pm0.03$    &    2.90   &                  & \cite{barish,allday} \\
      $\gamma$         & $6.68\pm0.48$    &    6.75   &                  & \cite{jaeger}     \\
      K$^0_S$          & $0.174\pm 0.011$ &    0.205  & 0.28/0.30$^{(a)}$& \cite{allday,jaeger}\\
      $\rho^0$         & $0.33 \pm0.06 $  &    0.39   & 0.49             & \cite{singer}           \\                      
      $\Lambda$        & $0.098\pm0.010$  &    0.124  & 0.27             & \cite{allday,jaeger}\\
      $\bar\Lambda$    & $0.014\pm0.004$  &    0.011  & 0.15             & \cite{allday,jaeger}\\ \hline                 
\multicolumn{5}{l} {$\sqrt{s}= 23.3$--$23.7 $~GeV ($\sigma_{inel}=32.21$~mb~\cite{amaldi})}             \\  \hline 
      Charged          & $9.24\pm 1.39$     &   8.61   &       & \cite{rossi}\\
      $\pi^0$          & $3.42\pm0.62$      &   3.79   & 0.17  & \cite{kafka}\\
      $\pi^+$          & $3.71\pm0.37$      &   3.67   & 0.19  & \cite{rossi}\\
      $\pi^-$          & $3.27\pm0.33$      &   3.14   & 0.17  & \cite{rossi}\\
      K$^0_S$          & $0.214\pm0.025$    &   0.250  & 0.28/0.30$^{(a)}$& \cite{lopinto}      \\
      K$^+$            & $0.337\pm0.051 $   &   0.344  & 0.33  & \cite{rossi}\\
      K$^-$            & $0.209\pm0.031 $   &   0.212  & 0.28  & \cite{rossi}\\
      K$^{*\pm}$       & $0.137\pm0.043$    &   0.215  & 0.60/0.67$^{(b)}$& \cite{lopinto}\\
      p                & $1.63\pm0.24 $     &   1.10   & 0.24  & \cite{rossi}\\
 $\bar{\mathrm{p}}$    & $0.085\pm0.013 $   &   0.063  & 0.16  & \cite{rossi}\\  
      $\Lambda$        & $0.112\pm 0.016$  &    0.131  & 0.27  & \cite{lopinto}      \\
      $\bar\Lambda$    & $0.020\pm0.004$   &    0.018  & 0.15  & \cite{lopinto}      \\  
      $\Sigma^{*\pm}$  & $0.017\pm0.012$   &    0.021  & 1.00  & \cite{lopinto}      \\  
$\bar{\Sigma}^{*\pm}$  & $0.014\pm0.011$   &    0.004  & 1.00  & \cite{lopinto}      \\ \hline 
\multicolumn{5}{l} {$\sqrt{s} = 26.0$ GeV ($\sigma_{inel}=32.8$~mb~\cite{bailly})}           \\  \hline 
      Charged          & $9.06\pm 0.09$    &    8.92  &                   & \cite{bailly} \\
   Neg. charged        & $3.53\pm0.05$     &    3.60  &                   & \cite{bailly} \\
      K$^0_S$          & $0.26\pm0.01$     &    0.26  & 0.28/0.30$^{(a)}$ & \cite{asai} \\
      $\Lambda$        & $0.12\pm 0.02$    &    0.13  & 0.26             & \cite{asai} \\
      $\bar\Lambda$    & $0.013\pm 0.004 $ &    0.020 & 0.15             & \cite{asai} \\ \hline 
\multicolumn{5}{l} {(a) Primary fraction of K$^0$ and $\bar{\mbox{K}}^0$, respectively.}  \\   
\multicolumn{5}{l} {(b) Primary fraction of K$^{*-}$ and K$^{*+}$, respectively.}  \\ \hline   
\end{tabular}
\end{center}
\caption{{\small  Average hadron production rates per inelastic
            pp collision at various centre-of-mass energies
           (excluding charge conjugates and antiparticles if not 
           indicated), compared with 
            the calculated values. The fraction of primary hadrons 
            obtained from the fit is also shown. For the calculated
            rates and fractions see discussions  in Section~3.}}
\label{pprate1}
\end{table}
                                                       
\newpage

\begin{table}[th]
\begin{center}
\begin{tabular}{lccccc} \hline   
 Hadron         &  Rate     &  Rate      & Prim. Fraction   & References & Prim. Fraction \\   
                & Measured  & Calculated & Calculated       &            & $\mbox{e}^{+}\mbox{e}^{-}$~\cite{pei}\\ \hline   
\multicolumn{5}{l} {$\sqrt{s}=27.4$--$27.6$~GeV ($\sigma_{inel}=32.8$~mb~\cite{LEBC})} \\   \hline 
    $\pi^0$            & $3.87\pm0.12$      &  3.82    & 0.17  & \cite{LEBC}    & 0.16 \\
    $\pi^+$            & $4.10\pm0.11$      &  3.70    & 0.19  & \cite{LEBC}    & 0.18 \\
    $\pi^-$            & $3.34\pm0.08$      &  3.17    & 0.17  & \cite{LEBC}    & 0.18 \\
    K$^0_S$            & $0.232\pm0.011 $   &  0.252   & 0.28/0.30$^{(a)}$& \cite{kichimi} & 0.25 \\
    K$^+$              & $0.331\pm0.016 $   &  0.346   & 0.33  & \cite{LEBC}    & 0.23 \\
    K$^-$              & $0.224\pm0.011 $   &  0.215   & 0.28  & \cite{LEBC}    & 0.23 \\
    $\eta$             & $0.30\pm0.02 $     &  0.33    & 0.28  & \cite{LEBC}    & 0.30 \\
    $\rho^0$           & $0.385\pm0.018 $   &  0.460   & 0.50  & \cite{LEBC}    & 0.47 \\
    $\rho^+$           & $0.552\pm0.082 $   &  0.490   & 0.54  & \cite{LEBC}    & 0.58 \\
    $\rho^-$           & $0.354\pm0.058 $   &  0.408   & 0.48  & \cite{LEBC}    & 0.58 \\
    K$^{*0}$           & $0.120\pm0.021 $   &  0.117   & 0.59  & \cite{LEBC}    & 0.49 \\
$\bar{\mbox{K}}^{*0}$  & $0.090\pm0.016 $   &  0.079   & 0.59  & \cite{LEBC}    & 0.49 \\
    K$^{*+}$           & $0.132\pm0.016 $   &  0.137   & 0.67  & \cite{LEBC}    & 0.50 \\
    K$^{*-}$           & $0.088\pm0.012 $   &  0.080   & 0.60  & \cite{LEBC}    & 0.50 \\
    $\omega$           & $0.391\pm0.024 $   &  0.421   & 0.52  & \cite{LEBC}    & 0.48 \\
    $\phi$             & $0.0189\pm0.0018 $ &  0.0175  & 1.00  & \cite{LEBC}    & 0.64 \\  
    f$_0(980)$         & $0.023\pm0.008 $   &  0.035   & 1.00  & \cite{LEBC}    & 0.99 \\
    f$_2(1270)$        & $0.092\pm0.012 $   &  0.090   & 0.77  & \cite{LEBC}    & 0.78 \\
K$^{*\pm}_{2}(1430)$   & $0.11\pm0.05 $     &  0.04    & 1.00  & \cite{kichimi} & 1.00 \\
      p                & $1.20\pm0.097 $    &  1.11    & 0.24  & \cite{LEBC}    & 0.12 \\
 $\bar{\mbox{p}}$      & $0.063\pm0.002 $   &  0.064   & 0.16  & \cite{LEBC}    & 0.12 \\
   $\Lambda$           & $0.125\pm0.008 $   &  0.131   & 0.27  & \cite{kichimi} & 0.12 \\
  $\bar\Lambda$        & $0.020\pm0.004 $   &  0.018   & 0.15  & \cite{kichimi} & 0.12 \\   
 $\Sigma^{+}$          & $0.048\pm0.015 $   &  0.039   & 0.48  & \cite{LEBC}    & 0.42 \\
 $\Sigma^{-}$          & $0.013\pm0.006 $   &  0.021   & 0.10  & \cite{LEBC}    & 0.42 \\
 $\Delta^{++}$         & $0.218\pm0.0031 $  &  0.214   & 0.83  & \cite{LEBC}    & 0.69 \\
 $\bar\Delta^{--}$     & $0.0128\pm0.0049$  &  0.0105  & 0.72  & \cite{LEBC}    & 0.69 \\
 $\Delta^{0}$          & $0.141\pm0.008  $  &  0.134   & 0.73  & \cite{LEBC}    & 0.69 \\
 $\bar\Delta^{0}$      & $0.0336\pm0.010 $  &  0.011   & 0.71  & \cite{LEBC}    & 0.69 \\
 $\Sigma^{*+}$         & $0.0204\pm0.0024 $ &  0.0193  & 1.00  & \cite{LEBC}    & 0.91 \\
 $\Sigma^{*-}$         & $0.0100\pm0.0018 $ &  0.0022  & 1.00  & \cite{LEBC}    & 0.91 \\
 $\bar\Sigma^{*\pm}$   & $0.0078\pm0.0025 $ &  0.0045  & 1.00  & \cite{LEBC}    & 0.91 \\
 $\Lambda(1520)$       & $0.0171\pm0.0031 $ &  0.0161  & 0.83  & \cite{LEBC}    & 0.71 \\ \hline
\multicolumn{5}{l} {(a) Primary fraction of K$^0$ and $\bar{\mbox{K}}^0$, respectively.}  \\ \hline  
\end{tabular}
\end{center}
\caption{{\small  Average hadron production rates per inelastic
            pp collision at $\sqrt{s}=27.4 - 27.6$~GeV
           (excluding charge conjugates and antiparticles if not 
           indicated), compared with 
            the calculated values. The fraction of primary hadrons 
            obtained from the fit is also shown. For the calculated
            rates and fractions see discussions  in Section~3.
            For comparison, the fraction of primary hadrons 
            obtained from the fit to $\mbox{e}^{+}\mbox{e}^{-}$ data
            at 91~GeV~\cite{pei} is also given in the last 
            column (the fraction values have been averaged for baryons 
            belonging to the same isospin multiplet).}}
\label{pprate2}
\end{table}

\newpage

\begin{table}[th]
\begin{center}
\begin{tabular}{lcccc} \hline   
 Hadron      &  Rate     &  Rate      & Primary Fraction & References  \\   
             &  Measured & Calculated & Calculated       &  \\ \hline   
\multicolumn{5}{l} {$\sqrt{s}=30.6 $~GeV } \\   \hline 
    Charged            & $10.07\pm 1.51$    &  9.87    &       & \cite{rossi}\\
    $\pi^+$            & $4.07\pm0.41$      &  4.21    & 0.20  & \cite{rossi}\\
    $\pi^-$            & $3.65\pm0.37$      &  3.68    & 0.18  & \cite{rossi}\\
    K$^+$              & $0.367\pm0.055 $   &  0.400   & 0.33  & \cite{rossi}\\
    K$^-$              & $0.244\pm0.037 $   &  0.269   & 0.29  & \cite{rossi}\\
      p                & $1.63\pm0.24 $     &  1.13    & 0.24  & \cite{rossi}\\
 $\bar{\mathrm{p}}$    & $0.108\pm0.016 $   &  0.091   & 0.16  & \cite{rossi}\\ \hline
\multicolumn{5}{l} {$\sqrt{s}=44.6 $~GeV } \\   \hline 
    Charged            & $10.99\pm 1.65$    &  10.82   &       & \cite{rossi}\\
    $\pi^+$            & $4.45\pm0.45$      &  4.62    & 0.20  & \cite{rossi}\\
    $\pi^-$            & $4.09\pm0.41$      &  4.09    & 0.18  & \cite{rossi}\\
    K$^+$              & $0.411\pm0.062 $   &  0.443   & 0.33  & \cite{rossi}\\
    K$^-$              & $0.286\pm0.043 $   &  0.312   & 0.29  & \cite{rossi}\\
      p                & $1.62\pm0.24 $     &  1.15    & 0.24  & \cite{rossi}\\
 $\bar{\mathrm{p}}$    & $0.132\pm0.020 $   &  0.112   & 0.16  & \cite{rossi}\\ \hline
\multicolumn{5}{l} {$\sqrt{s}=52.5$--$53.0$~GeV ($\sigma_{inel}=35.0$~mb~\cite{PDG})} \\   \hline 
    Charged            & $11.47\pm 1.72$    &  11.24   &       & \cite{rossi}\\
    $\pi^+$            & $4.68\pm0.47$      &  4.80    & 0.20  & \cite{rossi}\\
    $\pi^-$            & $4.29\pm0.43$      &  4.27    & 0.18  & \cite{rossi}\\
    K$^0_S$            & $0.329\pm0.071 $   &  0.359   & 0.29/0.31$^{(a)}$& \cite{drijard}\\
    K$^+$              & $0.430\pm0.065 $   &  0.462   & 0.33  & \cite{rossi}\\
    K$^-$              & $0.306\pm0.046 $   &  0.330   & 0.29  & \cite{rossi}\\
    $\rho^0$           & $0.63\pm0.14 $     &  0.62    & 0.51  & \cite{drijard}\\
$\bar{\mbox{K}}^{*0}$  & $0.123\pm0.028 $   &  0.123   & 0.60  & \cite{drijard}\\
    $\phi$             & $0.037\pm0.010 $   &  0.026   & 1.00  & \cite{drijard}\\  
    f$_2(1270)$$^{(b)}$ & $0.154\pm0.034 $  &  0.122   & 0.77  & \cite{drijard}\\
    f$_2(1270)$$^{(b)}$ & $0.075\pm0.007 $  &  0.122   & 0.77  & \cite{bohm}\\
K$^{*\pm}_{2}(1430)$$^{(b)}$ & $0.0044\pm0.0017 $ &  0.0264  & 1.00  & \cite{bohm}\\
$\bar{\mbox{K}}^{*\pm}_{2}(1430)$$^{(b)}$ & $0.028\pm0.007 $   &  0.0205  & 1.00  & \cite{drijard}\\
$\bar{\mbox{K}}^{*\pm}_{2}(1430)$$^{(b)}$ & $0.0031\pm0.0015 $ &  0.0205  & 1.00  & \cite{bohm}\\
      p                & $1.62\pm0.24 $     &  1.16    & 0.24  & \cite{rossi}\\
 $\bar{\mathrm{p}}$    & $0.144\pm0.022 $   &  0.121   & 0.16  & \cite{rossi}\\ \hline
\multicolumn{5}{l} {(a) Primary fraction of K$^0$ and $\bar{\mbox{K}}^0$, respectively.}  \\   
\multicolumn{5}{l} {(b) Not included in the fit.}                                        \\ \hline  
\end{tabular}
\end{center}
\caption{{\small  Average hadron production rates per inelastic
            pp collision at various centre-of-mass energies
           (excluding charge conjugates and antiparticles if not 
           indicated), compared with 
            the calculated values. The fraction of primary hadrons 
            obtained from the fit is also shown. For the calculated
            rates and fractions see discussions  in Section~3.}}
\label{pprate3}
\end{table}

\newpage

\begin{table}[th]
\begin{center}
\begin{tabular}{lcccl} \hline   
 Hadron      &  Rate     &  Rate      & Primary Fraction & References  \\   
             &  Measured & Calculated & Calculated       &  \\ \hline   
\multicolumn{5}{l} {$\sqrt{s}=200$~GeV } \\   \hline 
     Charged           & $21.4\pm 0.4$   &  21.4   &         & \cite{ansorge}$^{(a)}$\\
      K$^0_S$          & $0.75\pm 0.09$  &  0.78   & 0.31    & \cite{ansorge} \\                    
      n                & $0.75\pm 0.10$  &  0.73   & 0.19    & \cite{ansorge}$^{(b)}$\\                      
      $\Lambda$        & $0.23\pm0.06$   &  0.16   & 0.20    & \cite{ansorge} \\                                  
      $\Xi^-$          & $0.015\pm 0.015$&  0.007  & 0.50    & \cite{ansorge} \\  \hline 
\multicolumn{5}{l} {$\sqrt{s}=546$~GeV } \\   \hline 
     Charged           & $29.4\pm0.3$    &  29.4   &         & \cite{ansorge}$^{(a)}$\\
      K$^0_S$          & $1.12\pm 0.08$  &  1.12   &  0.31   & \cite{ansorge} \\
      $\Lambda$        & $0.265\pm0.055$ &  0.205  &  0.19   & \cite{ansorge} \\
      $\Xi^-$          & $0.05\pm0.015$  &  0.011  &  0.50   & \cite{ansorge} \\ \hline 
\multicolumn{5}{l} {$\sqrt{s}=900$~GeV } \\   \hline 
     Charged           & $35.6\pm0.9$    &  35.7   &         & \cite{ansorge}$^{(a)}$\\
      K$^0_S$          & $1.37\pm0.13$   &  1.38   & 0.31    & \cite{ansorge} \\
      n                & $1.0\pm0.2$     &  1.1    & 0.18    & \cite{ansorge}$^{(b)}$\\
      $\Lambda$        & $0.38\pm0.08$   &  0.24   & 0.18    & \cite{ansorge} \\
      $\Xi^-$          & $0.035\pm0.020$ &  0.014  & 0.50    & \cite{ansorge} \\ \hline 
\multicolumn{5}{l} {(a) The average charged multiplicity value quoted in this reference} \\
\multicolumn{5}{l} { ~~~~~~is increased by one to include the leading particles.}  \\   
\multicolumn{5}{l} {(b) The average production rate of the neutron quoted in this reference} \\
\multicolumn{5}{l} { ~~~~~~is increased by 0.5 to include the leading particles.}  \\ \hline  
\end{tabular}
\end{center}
\caption{{\small  Average hadron production rates per non-single-diffractive
            p$\bar{\mbox{p}}$ event at various centre-of-mass energies
           (excluding charge conjugates and antiparticles), compared with 
            the calculated values. The fraction of primary hadrons 
            obtained from the fit is also shown. For the calculated
            rates and fractions see discussions  in Section~3.}}
\label{pprate4}
\end{table}


\begin{table}[h]
\begin{center}
\begin{tabular}{lcccc} \hline   
$\sqrt{s}$(GeV) &  $C$      &  $f$     &  $x_{s}$    & $\chi^{2}/dof$   \\  \hline    
\multicolumn{5}{l} {pp} \\   \hline 
19.4--19.7   & $0.026\pm 0.004$   & 0.044 (fixed)  &  0.51 (fixed)  & 25/6  \\
23.3--23.7   & $0.041\pm 0.006$   & 0.044 (fixed)  &  0.51 (fixed)  & 16/13 \\
26.0         & $0.046\pm 0.007$   & 0.044 (fixed)  &  0.51 (fixed)  &  7/4  \\
27.4--27.6   & $0.042\pm 0.007$ & $ 0.044\pm 0.010$& $0.51\pm 0.22$ & 85/30 \\
30.6         & $0.060\pm 0.010$   & 0.044 (fixed)  &  0.51 (fixed)  &  6/6  \\
44.6         & $0.074\pm 0.014$   & 0.044 (fixed)  &  0.51 (fixed)  &  6/6  \\
52.5--53.0   & $0.080\pm 0.010$   & 0.044 (fixed)  &  0.51 (fixed)  &  7/10 \\  \hline 
\multicolumn{5}{l} {p$\bar{\mbox{p}}$} \\   \hline 
200          & $0.229\pm 0.035$   & 0.044 (fixed)  &  0.51 (fixed)  &  2/4  \\
546          & $0.347\pm 0.053$   & 0.044 (fixed)  &  0.51 (fixed)  &  8/3  \\
900          & $0.439\pm 0.066$   & 0.044 (fixed)  &  0.51 (fixed)  &  4/4  \\  \hline    
\end{tabular}
\end{center}
\caption{{\small  Results of the fit to pp and p$\bar{\mbox{p}}$ data obtained at
                  various centre-of-mass energies. The errors given in the Table
                  also include the uncertainty on the parameters
                  which are set to their central value in the fit.}}
\label{ppfit}
\end{table}

\newpage
\begin{figure}[th]
\vspace*{-1.8cm} 
\mbox{\epsfig{file=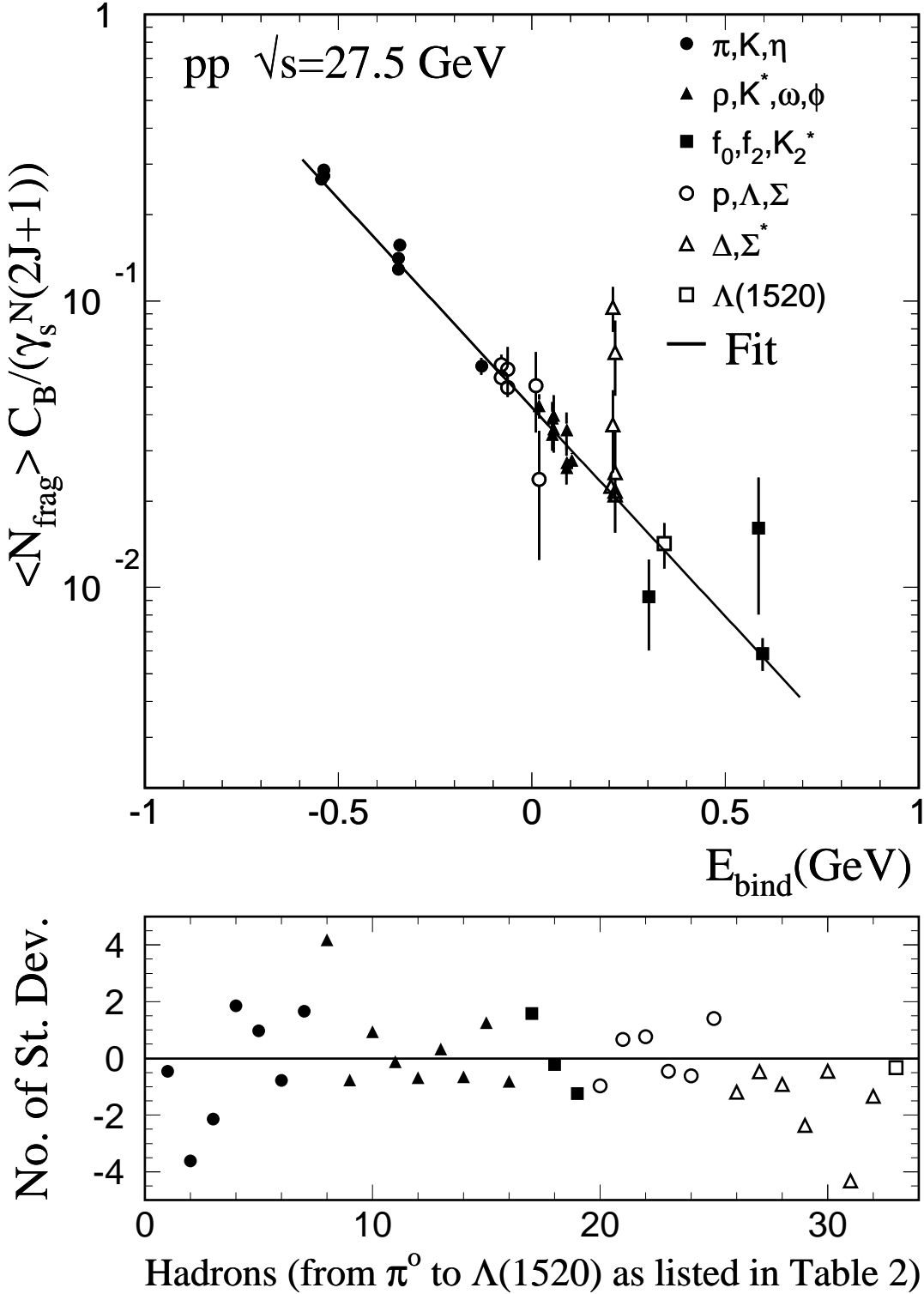,width=16cm}}
\vspace*{-2.5cm} 
\caption{{\small The upper plot shows the average 
hadron production rates of light-flavoured hadrons 
origi\-nating from fragmentation at $\sqrt{s}=27.4$--$27.6$~GeV in inelastic pp collisions 
(measured value $\times$ fraction of hadrons originating from fragmentation
as determined by the fit), multiplied by the factor 
$C_{B}/[\gamma_{s}^{N_{s}}(2J+1)]$ (see Eq.~(\ref{eqrate})),
as a function of the binding energy of hadrons. The fit results are shown as the line.
The lower plot shows the difference between the predicted and measured rates in terms
of the number of standard deviations.
}}
\end{figure}

\end{document}